\documentclass[preprint]{elsarticle}

\usepackage{lineno}
\modulolinenumbers[5]

\begin{document}

\begin{frontmatter}

\title{Development of optical resonant cavities for laser-Compton scattering}

\author[KEK]{T. Akagi}

\author[KEK]{S. Araki}


\author[KEK]{Y. Honda}

\author[KEK]{A. Kosuge}


\author[KEK]{T. Omori}

\author[KEK]{H. Shimizu}

\author[KEK]{N. Terunuma}

\author[KEK]{J. Urakawa}


\author[Hiroshima]{T. Takahashi\corref{mycorrespondingauthor}}
\cortext[mycorrespondingauthor]{Corresponding author}
\ead{tohru-takahashi@hiroshima-u.ac.jp}

\author[Hiroshima]{R. Tanaka}

\author[Tohoku]{Y. Uesugi}

\author[Hiroshima]{H. Yoshitama}

\author[Waseda]{Y. Hosaka}

\author[Waseda2]{K. Sakaue}

\author[Waseda]{M. Washio}

\address[KEK]{High Energy Accelerator Research Organization (KEK), 1-1 Oho, Tsukuba, Ibaraki 305-0801, Japan}
\address[Hiroshima]{AdSM Hiroshima University, 1-3-1 Kagamiyama, Higashi Hiroshima, 739-8530, Japan}
\address[Tohoku]{Institute of Multidisciplinary Research for Advanced Materials, Tohoku University, Katahira 2-1-1, Aoba-ku, Sendai 980-8577, Japan}
\address[Waseda]{Research Institute for Science and Engineering, Waseda University, 3-4-1 Okubo, Shinjuku-ku, 169-8555, Japan}
\address[Waseda2]{Waseda Institute for Advanced Study, Waseda University, 3-4-1 Okubo, Shinjuku-ku, 169-8555, Japan}

\begin{abstract}
We have been developing optical resonant cavities for
 laser-Compton scattering experiment at the Accelerator Test Facility in KEK.
The main subject of the R\&D is to increase
laser pulse energy by coherently accumulating the pulses in an optical resonant cavity.
We report previous results, current status and future prospects, including a new idea 
of an optical resonant cavity.
\end{abstract}

\begin{keyword}
laser-Compton scattering, 
optical resonant cavity, 
self-resonating cavity 
\end{keyword}

\end{frontmatter}


\section{Introduction}

Laser-Compton scattering is an attractive method to generate photons 
in an energy range from x to $\gamma$ rays.
The advantages of of this scheme are; 
the required electron energy is lower than those of synchrotron radiation, 
polarized photons can be produced and easily controlled by the incident laser 
beam polarization.
Thanks to these advantages, 
this scheme can be applied to a wide area of scientific and industrial fields such as
material science, medical science, particle-nuclear physics,
and therefore, its characteristics and possible applications  
have been discussed by many authors 
\cite{Krafft, Sun, Brown, Petrillo1, Petrillo2, Serafini, CAIN, Telnov, Omori, Chiche, Bech, Sakaue}.

One of the issues of the laser-Compton scheme is
 to obtain sufficient photon intensity for various applications.
An optical resonant cavity is an idea to provide high power and high repetition 
laser pulses by coherently accumulating them in the cavity, 
which allows us to construct a laser-Compton facility without a large scale 
laser  system.
For the coherent superposition of laser pulses, however, 
the length of the optical path in the cavity should 
be multiples of the wavelength with high precision, that 
reaches O (10) pm for the finesse of ${10^3} \sim {10^4}$.
Therefore, controlling  the optical path length is 
a key issue to realize the laser system with the high finesse optical cavity. 

We have developed optical cavities and performed a series of 
experiment to generate photon beams by laser-Compton scattering at 
the Accelerator Test Facility (ATF) in KEK, Japan, 
by the collaboration of KEK, Waseda University and Hiroshima University 
with a close information exchange with a group of LAL Orsay, France.
The initial results were published in \cite{Shimizu, Miyoshi}.
At the latest experiment, we successfully kept the optical path length of the cavity with the order of 10 pm and 
enhanced the laser pulse intensity to 1230 times in the cavity \cite{Akagi}.

 For  more power enhancement in the cavity, one order higher for example, 
the required precision  will reach O(1) pm.
It could be achieved with a sophisticated optical and feedback system, however, 
is technically challenging to develop such system.
 As an alternative of precise control of the optical path length by active feedback system, 
we proposed an idea of self-resonating cavity, in which the optical system composed of 
a laser amplifier and an optical cavity keeps resonance condition by itself \cite{Honda-free}.

In this article,  we report a status and prospect of our effort to develop optical cavities
for laser-Compton scattering, including a self-resonating cavity.

\section{4 Mirror Optical Cavity in the KEK ATF}

\begin{figure}[!h]
\centering\includegraphics[width=3.5in]{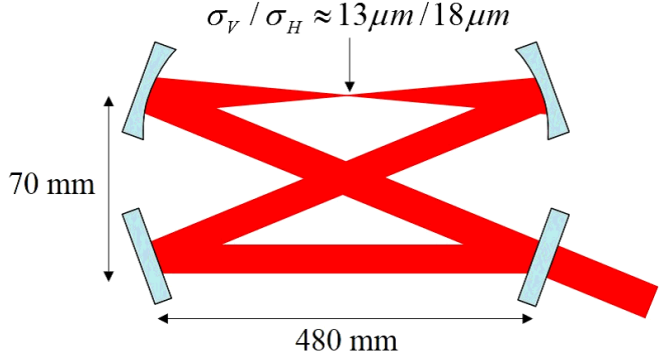}
\caption{A schematic of 4 mirror cavity. The cavity consists of two flat and two concave mirrors.
   The injected laser wave is focused at the center of the two concave mirrors, where the electron bunches 
    are passing through.The vertical and horizontal laser spot sized at the focal point are shown in the figure. }
\label{4mcavity}
\end{figure}

A schematic of an optical cavity installed in the ATF is shown in figure \ref{4mcavity}.
The cavity is a 4 mirror ring cavity with the optical path length of about 1.68 m.
A discussion about various configurations of the 4 mirrors cavities is find in \cite{Zomer}.
Comparing with a simple two mirror cavity, the 4 mirror cavity has advantages of 
stability against miss-alignment of the mirrors. particularly for 
an operation with a tight laser beam focusing \cite{Akagi}.  
The main parameters of the ATF and laser pulses used in the experiments are summarized in 
table \ref{parameters}.

\begin{table}[htb]
\caption {Main parameters of the election bunches  and laser pulses} 
\centering\begin{tabular}{|l| l| } \hline
\multicolumn{2} {|l|}{Electron}  \\ \hline 
Energy &  1.28 GeV \\ \hline
Energy Spread  $\delta p/p$ & $5.43\times 10^{-4}$ \\ \hline
Beam Intensity & $1\times 10^{10}$/bunch  \\ \hline
Bunch Spacing & 5.6 ns \\ \hline
Circumference & 138.6 m \\ \hline
Emittance ($\epsilon_x /\epsilon_y$) & $1.1\times 10^{-9}/12\times 10^{-12}$  \\ \hline
Revolution & 2.16 MHz \\ \hline \hline
 \multicolumn{2}{|l|}{Laser} \\ \hline 
Wavelength & 1064 nm \\ \hline
Repetition frequency & 357 MHz \\ \hline
Pulse width & 11 ps (FWHM)\\ \hline
Power & 10 W (28nJ/pulse) \\ \hline
  \end{tabular}
\label{parameters}
\end{table}

A unique feature of the cavity is that the 4 mirrors of the cavity are arranged 
in a way to make a twisted optical path for the circulating laser pulses. 
It causes a geometric phase for a laser wave state in the cavity, and as a result,  
the cavity only resonates with left- or right- circularly polarized state separately at 
different optical path length.
This feature is useful for generating circularly polarized photon via laser-Compton scattering 
and also applicable  to a feedback control scheme with a simple setup \cite{Honda-fb}.

We conducted the laser-Compton experiments at the KEK ATF, 
an electron storage ring which was constructed to develop technique for
an ultra-low emittance electron beam required for future linear colliders \cite{ATF}.

\begin{figure}[!h]
\centering\includegraphics[width=3.5in]{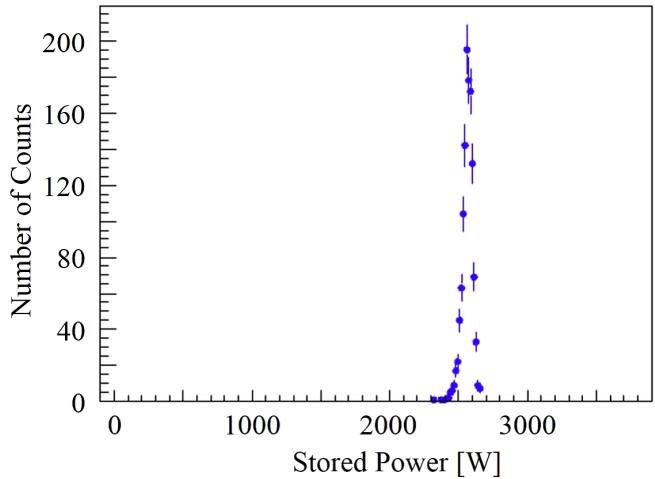}
\caption{Stored power in the optical resonant cavity during the photon generation experiment at the ATF.
The average power in the cavity was 2.6 kW and its fluctuation was 38 W. }
\label{storage}
\end{figure}

Figure \ref{storage} shows the stored power in the cavity during the operation in 
the ATF ring. 
The average power was observed to be 2.6 kW.
Since the input power to the cavity was 2.1 W, 
the stored power was enhanced by a factor of 1230. 
From the fluctuation of the stored power, 38 W, 
the precision of the optical path control was 
calculated to be  $\delta L_{cavity} \approx 16 $ pm.
We also succeeded to generate $\gamma$ rays at the rate of $2.7 \times 10^8/s$ with their average energy of 24 MeV


It must be mentioned, while not discussed in this article,
another 4 mirror optical cavity was installed by the French group
in the ATF and successfully demonstrated photon generation of 
$(3.5 \pm 0.3) \times 10^8$ photons per second with an average power in the cavity of 
35 kW \cite{Chaikovska, Bonis}.
The cavity equipped an automated mirror alignment system with Pound-Drever-Hall stabilizing technique, 
while the cavity described in this article adopted relatively simple mechanical configuration and the polarization based feedback system.
The average photon yield depended on the electron beam conditions during the experiments, which eventually 
gave us a similar number of photons per second. 
The photon yields normalized by the electron intensity for both experiments were consistent with the laser power and
the French cavity showed approximately 400 photons per laser-election pulse interaction, which is the world record at the time of 
the publication \cite{Chaikovska}.

\section{Self-resonating Optical cavity}

\begin{figure}[!h]
\centering\includegraphics[width=3.5in]{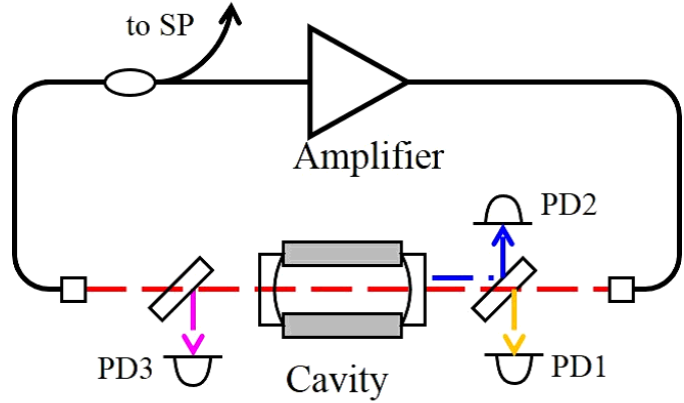}
\caption{A schematic of the optical setup of self resonating cavity.
The laser power was monitored by Photo-Detectors (PDs) and the 
spectrum of the circulation laser waves was monitored by a Spectrum Analyzer(SP).}
\label{self-setup}
\end{figure}

The obtained photon yields of $O(10^8)/s$, described in the previous section, are still not sufficient for 
realistic applications, thus we need to increase intensity of both electron and laser beams.   
In terms of the laser power in the optical cavities, the required precision of the optical path control will  reach 
order of picometers or less.
To avoid the technical difficulty, an idea of the self-resonating cavity was proposed \cite{Honda-free}.
A configuration of the self-resonating cavity is illustrated in figure \ref{self-setup}, 
which  forms an optical loop with a laser amplifier and an optical cavity.
At the beginning, a laser wave which satisfies the resonance condition, 
$L_{cavity} = n \lambda_{laser}/2$,  
circulates in the optical loop and acts as a seed of the laser oscillation.
When the gain of the amplifier becomes sufficient to compensate over all power loss 
in the loop, the laser oscillation starts spontaneously with a wavelength which satisfies
the resonance condition.
The length of the cavity, $L_{cavity}$, and therefore the resonance wavelength, 
fluctuates due to environmental disturbances, however,
the system, in principle, keeps the  laser oscillation  
with a new wavelength chosen by the cavity.
  We demonstrated the idea with a two mirror Fabry Perrot cavity\cite{Uesugi}. 
The reflectivity of the mirrors are 99.999\% and the intrinsic finesse of the cavity was
 independently measured to be $646,000 \pm 3,000$.

\begin{figure}[hbt]
\centering\includegraphics[width=3.in]{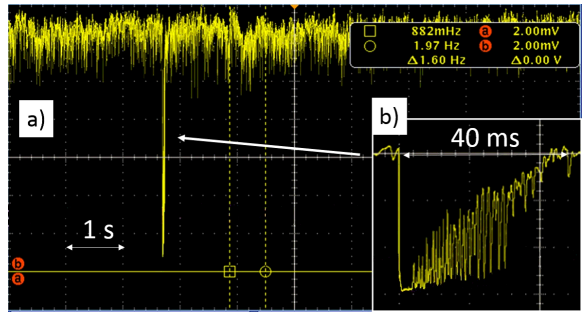}
\caption{A demonstration of the stability of the self-resonating cavity. The line is a temporal behavior
of the laser power in the cavity monitored by PD3 in figure \ref{self-setup}.
What shown in b) is a magnification of the behavior at the power recovery. 
See text for the description of a sharp drop the power and its recovery.}
\label{drop}
\end{figure}

As the first trial, we operated the system with continuous wave laser with the wavelength around 
1047 nm.
We observed a highly stable operation with the power fluctuation of $\sigma = 1.7 $\% for more than 
two hours.
Figure \ref{drop} shows a demonstration of the stability of the self-resonating cavity. 
During a stable operation of the cavity, we hit the body of the cavity with a hammer.
A sharp drop of the stored power was observed,  but was automatically restored in 
about 40 ms, which was consistent with a damping time of the mechanical vibration 
of the body of the cavity.
The effective finesse, which is a measure of the performance of the cavity including
the stability of self-resonating mechanism, was measured to be 
$394,000 \pm 10,000$.
The stored power in the cavity, which was calculated from the transmitted power with the transmittance of the mirror, 
was $2.52 \pm 0.13 $ kW, showing a power enhancement factor of approximately 187,000.

\begin{figure}[!h]
\centering\includegraphics[width=3.5in]{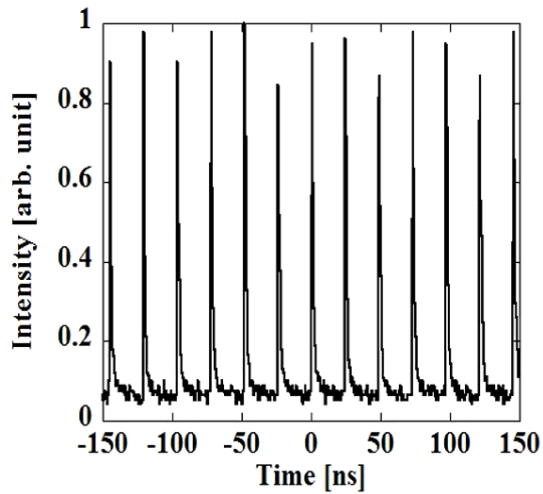}
\caption{An observed temporal waveform during a mode locked oscillation of the self-resonating cavity.  }
\label{self-modelock}
\end{figure}

Toward application of this idea to the laser-Compton scattering, the mode locked pulsed oscillation is
necessary. 
It requires a mechanism to make pulses in the system such as a nonlinear polarization rotation optics.
In addition, mode locked oscillation requires a relation between the entire optical path and the cavity length as 
$L_{Loop} = n L_{cavity}$, where $L_{Loop}$ and $L_{cavity}$ are the optical path length of the entire loop and the 
length of the cavity embedded in the optical loop.
We succeeded a proof of the principle experiment of a mode locked oscillation with low finesse (about 220) cavity
as shown in figure \ref{self-modelock}.
The details of the mode locked operation can be found in \cite{Hosaka-preprint} and 
the R\&D toward a stable and high power storage is now in progress.

\section{Conclusion}

A series of experiments has been performed at the ATF electron ring in KEK and demonstrated 
photon generation by laser-Compton scattering.

A new cavity based on the self-resonating mechanism has been developed and 
we successfully demonstrated a stable operation with continuous wave.
An effort is in progress toward realization of stable mode lock operation 
for the laser-Compton experiment.

\section*{Acknowledgment}

The authors would like to thank the members of the ATF group and
collaborators of LAL Orsay, France. 
This research has been supported by Quantum Beam Technology Program 
of Japanese Mext and  
 also supported by JSPS KAKENHI Grant numbers 23226020,  24654076
and 25246039




\begin{thebibliography}{9}

\bibitem{Krafft}
G. A. Krafft and G. Priebe, Rev. Accel. Sci. Technol. Vol. 3 (2010) 147-163

\bibitem{Sun}
 C. Sun and Y. K. Wu, Phys. Rev. ST Accel. Beams 14, 044701 (2011)

\bibitem{Brown}
W. J. Brown and F. V. Hartemann, Phys. Rev. ST Accel. Beams 7, 060703 (2004)

\bibitem{Petrillo1}
V. Petrillo et al., Nucl. Instr. Meth. Phys. Res. A 693 (2012) 109

\bibitem{Petrillo2}
V. Petrillo et al., Phys. Rev. ST Accel. Beams 18 (2015)

\bibitem{Serafini}
 L. Serafini et al, EPJ Web of Conferences 117 (2016) 05002

\bibitem{CAIN}
P. Chen, et.al,  Nulc. Instr. and Meth. A397 458-464 (1997), 
https://ilc.kek.jp/~yokoya/CAIN/cain235/
 
\bibitem{Telnov}
B. Badelek, et.al, Int. J. of Mod. Phys. A19 5097 (2004)

\bibitem{Omori}
T. Omori, et. al, Phys. Rev. Lett. 96, 114801 (2006)

\bibitem{Chiche}
R. Chiche, et. al, Appl. Opt 52 8377 (2013) 

\bibitem{Bech}
M. Bech, et. al,  J. Synchrotron Rad. 16, 43 (2009).

\bibitem{Sakaue}
K. Sakaue, et. al,  Rev. Sci. Intr. Vol. 80. No. 12, 123304 (2009).

\bibitem{Shimizu}
H. Simizu et. al, J. of Phys. Sco. of Japan. {\bf 78-7}, 074501 (2009).

\bibitem{Miyoshi}
S. Miyoshi et. al, Nucl. Instr. and Meth. A {\bf 623}, 576 (2010).

\bibitem{Akagi}
T. Akagi et. al, Nucl. Instr. and Meth. A {\bf 724}, 63 (2013).

\bibitem{Zomer}
F. Zomer, et. al,  Appl. Opt. Vol. 48, No. 35, 6651 (2009). 

\bibitem{Honda-fb}
Y. Honda, et.al, Opt. Comm. Vol. 282. No. 15, 3108 (2009)
T. Akagi, et. al,  Rev. Sci. Instrum. 86, 043303 (2015).

\bibitem{ATF}
J. Urakawa et al. (ATF Collaboration), http://atf.kek.jp/
collab/ap/.


\bibitem{Borzsonyi}
A. Borzsonyi, et.al, Appl. Opt. 52, 8377  (2013).

\bibitem{Chaikovska}
  I. Chaikovska et al, Scientific Reports 6 (2016)

\bibitem{Bonis}
 J. Bonis et al, J. of Inst 7 (2012) P01017

\bibitem{Honda-free}
Y. Honda, et. al. Proc. 7th Annual Meeting of PASJ, 1102 (2010), In Japanese. 

\bibitem{Uesugi}
Y. Uesugi, et. al, APL Photonics 1, 026103 (2016).

\bibitem{Hosaka-preprint}
Y. Hosaka, et. al, arXiv:1610.03141

\end{thebibliography}

\end{document}